\documentclass[aps,preprint]{revtex4}%
\usepackage{amsmath}
\usepackage{amsfonts}
\usepackage{amssymb}
\usepackage{graphicx}%
\providecommand{\U}[1]{\protect\rule{.1in}{.1in}}

\begin{document}
\title{Networks of dissipative quantum harmonic oscillators: a general treatment}
\author{M. A. de Ponte$^{1}$, S. S. Mizrahi$^{1}$, and M. H. Y. Moussa$^{2}$.}
\affiliation{$^{1}$Departamento de F\'{\i}sica, Universidade Federal de S\~{a}o Carlos,
Caixa Postal 676, S\~{a}o Carlos, 13565-905, S\~{a}o Paulo,\textit{ }Brazil}
\affiliation{$^{2}$ Instituto de F\'{\i}sica de S\~{a}o Carlos, Universidade de S\~{a}o
Paulo, Caixa Postal 369, 13560-590 S\~{a}o Carlos, SP, Brazil }

\begin{abstract}
In this work we present a general treatment of a bosonic dissipative network:
a chain of coupled dissipative harmonic oscillators whichever its topology,
i.e., whichever the\textit{\ way} the oscillators are coupled together,
the\textit{\ strenght} of their couplings and their \textit{natural
frequencies}. Starting with a general more realistic scenario where each
oscillator is coupled to its own reservoir, we also discuss the case where all
the network oscillators are coupled to a common reservoir. We obtain the
master equation governing the dynamic of the network states and the associated
evolution equation of the Glauber-Sudarshan $P$-function. With these
instruments we breafly show how to analyse the decoherence and the evolution
of the linear entropy of general states of the network. We also show how to
obtain the master equation for the case of distinct reservoirs from that of a
common one.

\end{abstract}

\maketitle

\section{Introduction}

Over the last few years interest has grown for better understanding the
phenomena of coherence and decoherence dynamics in quantum networks,
especially in connection to the protocols for quantum-state transferring and
quantum-state protection for information processing. Beyond the quest for
conditions that weaken the system-reservoir coupling \cite{Landauer,Unruh},
the search for mechanisms to bypass decoherence started with
quantum-error-correction codes \cite{QECC} and goes through the program of
engineering reservoirs \cite{Poyatos}. Moreover, in a closer contact with
quantum networks, the investigation of collective decoherence resulted on what
has been called a decoherence-free subspace \cite{ZR,DFS,Lidar}. Interestingly
enough, while, in general, the protocols for quantum error-correcting codes
presuppose that quantum systems decohere independently, the decoherence-free
subspace is generated by distinct quantum systems coupled to a common reservoir.

Regarding state transfer, the controlled coherent transport with splitting of
atomic wave packets \cite{Bloch}, and the evolution of macroscopically
entangled state \cite{Polkovnikov} have been analyzed within the context of
optical lattices. The dynamics of Bose-Einstein condensates in a
one-dimensional optical lattice is also investigated \cite{Fromhold}, and a
class of spin networks have been proposed for perfect state transfer of any
quantum state in a fixed period of time \cite{Landahl}. In the context of
systems of coupled harmonic oscillators (HOs), which we focus in the present
work, the dynamics and manipulation of entanglement was analyzed in Ref.
\cite{Plenio}. Evidently, the pressure for the implementation of logical
operations with an increasingly larger number of quantum systems will
decisively drive on the quest for controlled coherent transport in quantum networks.

Concerned with a simple network of two coupled resonators, Raimond \textit{et
al}. \cite{Raimond} have presented the blueprint of an experiment in which the
decoherence of a mesoscopic superposition of radiation states becomes a
reversible process. A theoretical model of the proposal in Ref. \cite{Raimond}
is given in Ref. \cite{Nemes}, where the coupling of the resonators to their
environments is taken into account when the reversibility of coherence loss is
analyzed. In Ref. \cite{Zoubi}, the authors assume that only one of the
resonators in Refs. \cite{Raimond,Nemes} is interacting with a reservoir to
derive a master equation in the case where the resonators are strongly
coupled. It is shown that the relaxation term is not simply the standard one,
obtained by neglecting the interaction between the cavities, i.e., dissipation
is not additive for strongly coupled systems. Finally, in Ref. \cite{Mickel}
both resonators are considered to be lossy, as in Refs. \cite{Nemes,Raimond},
and the regime of strongly coupled cavities is also analyzed, as done in Ref.
\cite{Zoubi}. A detailed analyses of the coherence and decoherence dynamics of
quantum states is presented in \cite{Mickel}, regarding this network of two
coupled resonators, including a study of the correlation between the fields in
both resonators through the excess entropy. The phenomena of
electromagnetically induced transparency and dynamical Stark effect are also
analyzed in a network of two coupled dissipative resonators \cite{Mickel-EPL}.

In the context of complex networks, composed by a large number of subsystems,
in Refs. \cite{Mickel1,Mickel2} the authors present a detailed treatment of
the coherence and decoherence dynamics in arrays of coupled dissipative
resonators. In Ref. \cite{Mickel1} a symmetric network of $N$ interacting
resonators are considered, where each oscillator interacts with each other,
apart from its own reservoir. A different topology is analyzed in Ref.
\cite{Mickel2}, where a central oscillator is assumed to interact with the
remaining $N-1$ peripheral and noninteracting oscillators. In both topologies,
the decoherence process is analyzed by focusing on a single resonator which,
apart from interacting with its own reservoir, also interacts with the
remaining $N-1$ coupled resonators plus their respective reservoirs.
Considering all resonators with the same natural frequency $\omega_{0}$ and
all couplings with the same strength $\lambda$, master equations are derived
for both weak ($\lambda\ll\omega_{0}$) and strong ($\lambda\approx\omega_{0}$)
coupling regimes. From such development, a detailed analyzes of the emergence
of relaxation- and decoherence-free subspaces in networks of weakly and
strongly coupled resonators is presented in Ref. \cite{Mickel3}. The main
result in Ref. \cite{Mickel3} is that both subspaces are generated when all
the resonators couple with the same group of reservoir modes, thus building up
a correlation (among these modes), which has the potential to shield
particular network states against relaxation and/or decoherence.

It is worth noting that recent results regarding entanglement and nonclassical
effect in collective two-atom systems \cite{Ficek}, retains some resemblance
with those discussed above for networks of coupled resonators. Beyond the
entanglement dynamics which is a crucial but recurrent ingredient of any
network, the collective damping effects coming from two-atom systems
\cite{Ficek} can be directly identified with those in a network of dissipative
oscillators \cite{Mickel,Mickel1,Mickel2,Mickel3}. Such collective damping
effect are certainly in the basis of the nonadditivity of decoherence rates
observed in the network of dissipative oscillators
\cite{Mickel,Mickel1,Mickel2,Mickel3} as well as in superconducting qubits
\cite{Brito}.

Since in Refs. \cite{Mickel1,Mickel2} two different topologies are analyzed
independently, for the particular case where all resonators have the same
natural frequency $\omega_{0}$ and all couplings have the same strength
$\lambda$, in this work we present a unified approach for treating a bosonic
dissipative network. Such approach holds for whichever the topology of the
network. i.e., for whichever $i)$ the way the resonators are coupled among
them, $ii)$ their coupling strengths and $iii)$ natural frequencies.

In the Section II, towards the derivation of the master equation governing the
evolution of the network, we present our model. Considering first a
nondissipative network, we show how to derive particular topologies from the
general case of a symmetric network where each oscillator interacts with each
other. In Section III, the evolution equation of the Glauber-Sudarshan
P-function is obtained as a c-number map of the master equation in operator
form. Thus, in the context of dissipative networks, we show how to derive
particular topologies from a general symmetric dissipative network where each
oscillator is coupled to its respective reservoir apart from interacting with
each other. Solutions in terms of the Glauber-Sudarshan P-function, for
general initial states of the network, are given in Section IV together with a
brief analysis of decoherence and the linear entropy. Finally, the concluding
remarks are presented in Section VI.

\section{General treatment of a bosonic network}

\subsection{The model}

Setting from here on that the indexes $m,m^{\prime},n,$ and $n^{\prime}$run
from $1$\ to $N$, we start considering a general Hamiltonian for a bosonic
network, $H=H_{S}+H_{R}+H_{I}$, involving a network of $N$ coupled
oscillators
\begin{equation}
H_{S}=\hbar\sum_{m}\omega_{m}a_{m}^{\dag}a_{m}+\frac{\hbar}{2}\sum_{m\neq
n}\lambda_{mn}\left(  a_{m}^{\dag}a_{n}+a_{m}a_{n}^{\dag}\right)  \text{,}
\label{1}%
\end{equation}
$N$ distinct reservoirs, composed by a set of $k=1,\ldots,\infty$ modes,%
\begin{equation}
H_{R}=\hbar\sum_{m}\sum_{k}\omega_{mk}b_{mk}^{\dag}b_{mk}\text{,} \label{2}%
\end{equation}
and the coupling between the network oscillators and their respective
reservoirs%
\begin{equation}
H_{I}=\hbar\sum_{m}\sum_{k}V_{mk}\left(  b_{mk}^{\dag}a_{m}+b_{mk}a_{m}^{\dag
}\right)  \text{.} \label{3}%
\end{equation}
$b_{mk}^{\dagger}$ ($b_{mk}$) is the creation (annihilation) operator for the
$k$th bath mode $\omega_{mk}$ coupled to the $m$th network oscillator
$\omega_{m}$ whose creation (annihilation) operator reads $a_{m}^{\dagger}$
($a_{m}$). The coupling strengths between the oscillators are given by the set
$\left\{  \lambda_{mn}\right\}  $, while those between the oscillators and
their reservoirs by $\left\{  V_{mk}\right\}  $. Before addressing the
dissipative process through Hamiltonians (\ref{2}) and (\ref{3}), we focus
first on Hamiltonian $H_{S}$ to show how to derive different topologies of a
nondissipative network of coupled harmonic oscillators. Rewriting $H_{S}$ in a
matrix form $H_{S}=\hbar\sum_{m,n}a_{m}^{\dag}\mathcal{H}_{mn}a_{n}$, its
elements are given by%
\begin{equation}
\mathcal{H}_{mn}=\left\{
\begin{array}
[c]{ccc}%
\omega_{m} & \text{if} & m=n\\
\lambda_{mn} & \text{if} & m\neq n
\end{array}
\right.  \text{,} \label{4}%
\end{equation}
whose values characterize whichever the network topology: the way the
oscillators are coupled together, the set of coupling strengths $\left\{
\lambda_{mn}\right\}  $, and their natural frequencies $\left\{  \omega
_{m}\right\}  $.

\subsection{From the general matrix $\mathcal{H}$ to particular nondissipative
topologies}

To illustrate the procedure to construct particular nondissipative topologies
we consider four different cases: the $i)$ symmetric, $ii)$ central, $iii)$
circular, and $iv)$ linear networks. For the case of a $i)$ symmetric ($sym$)
network, sketched in Fig. 1(a), all the oscillators are coupled together, with
all matrix elements of $\mathcal{H}$ being not null%
\begin{equation}
\mathcal{H}_{sym}=\left(
\begin{array}
[c]{ccccc}%
\omega_{1} & \lambda_{12} & \lambda_{13} & \cdots & \lambda_{1N}\\
\lambda_{12} & \omega_{2} & \lambda_{23} & \cdots & \lambda_{2N}\\
\lambda_{13} & \lambda_{23} & \omega_{3} & \cdots & \lambda_{3N}\\
\vdots & \vdots & \vdots & \ddots & \vdots\\
\lambda_{1N} & \lambda_{2N} & \lambda_{3N} & \cdots & \omega_{N}%
\end{array}
\right)  \text{.} \label{Sym}%
\end{equation}
In a $ii)$ central ($cent$) network, sketched in Fig. 1(b), only one selected
oscillator, the central one, is assumed to interact with the remaining $N-1$
noninteracting peripheral oscillators. Labeling the central oscillator by $1$,
with the peripherals running from $2$ to $N$, the matrix $\mathcal{H}$ has the
form%
\begin{equation}
\mathcal{H}_{cent}=\left(
\begin{array}
[c]{ccccc}%
\omega_{1} & \lambda_{12} & \lambda_{13} & \cdots & \lambda_{1N}\\
\lambda_{12} & \omega_{2} & 0 & \cdots & 0\\
\lambda_{13} & 0 & \omega_{3} & \cdots & 0\\
\vdots & \vdots & \vdots & \ddots & \vdots\\
\lambda_{1N} & 0 & 0 & \cdots & \omega_{N}%
\end{array}
\right)  \text{,} \label{Cent}%
\end{equation}
where only in first column and row all the nondiagonal elements are not null.
As depicted in Fig. 1 (c), in a $iii)$ circular ($circ$) network the $k$th
oscillator is coupled to the $\left(  k\pm1\right)  $th oscillators, with the
additional condition that the $N$th oscillator be coupled to the first one.
The matrix $\mathcal{H}$ is given by%
\begin{equation}
\mathcal{H}_{circ}=\left(
\begin{array}
[c]{ccccc}%
\omega_{1} & \lambda_{12} & 0 & \cdots & \lambda_{1N}\\
\lambda_{12} & \omega_{2} & \lambda_{23} & \cdots & 0\\
0 & \lambda_{23} & \omega_{3} & \cdots & 0\\
\vdots & \vdots & \vdots & \ddots & \vdots\\
\lambda_{1N} & 0 & 0 & \cdots & \omega_{N}%
\end{array}
\right)  \text{.} \label{Circ}%
\end{equation}
Finally, the $iv)$ linear ($lin$) network follows directly from the circular
one by turning off the coupling between the first and the $N$th oscillators.
The matrix $\mathcal{H}$ obtained for this case has the three-diagonal form%
\begin{equation}
\mathcal{H}_{lin}=\left(
\begin{array}
[c]{ccccc}%
\omega_{1} & \lambda_{12} & 0 & \cdots & 0\\
\lambda_{12} & \omega_{2} & \lambda_{23} & \cdots & 0\\
0 & \lambda_{23} & \omega_{3} & \cdots & 0\\
\vdots & \vdots & \vdots & \ddots & \vdots\\
0 & 0 & 0 & \cdots & \omega_{N}%
\end{array}
\right)  \text{.} \label{Lin}%
\end{equation}

Next, we treat the general situation of the dissipative network, described by
Hamiltonian $H=H_{S}+H_{R}+H_{I}$, through the standard perturbative approach
in the system-bath coupling strengths $\left\{  V_{mk}\right\}  $. Since all
the particular topologies follow from the case of a symmetric network,
choosing appropriately the elements of $\mathcal{H}$, we shall obtain the
reduced density operator of the coupled oscillators from this general topology.

\subsection{The master equation of a bosonic dissipative network -- Direct and
indirect dissipative channels}

To obtain the master equation of the network we first diagonalize the
Hamiltonian $\mathcal{H}$ through a canonical transformation%
\begin{equation}
A_{m}=\sum_{n}C_{mn}a_{n}\text{,} \label{5}%
\end{equation}
where the coefficients of the $m$th line of matrix $\mathbf{C}$ define the
eigenvectors associated to the eigenvalues $\varpi_{m}$ of matrix (\ref{Sym}).
With $\mathbf{C}$ being an orthogonal matrix, in that $\mathbf{C}%
^{T}=\mathbf{C}^{-1}$, it follows the commutation relations $\left[
A_{m},A_{n}^{\dag}\right]  =\delta_{mn}$ and $\left[  A_{m},A_{n}\right]  =0$,
enabling the Hamiltonian $H$ to be rewritten as $\widetilde{H}=H_{0}+V$, with
$a_{m}=\sum_{n}A_{n}C_{nm}$ and
\begin{subequations}
\label{6}%
\begin{align}
H_{0}  &  =\hbar\sum_{m}\varpi_{m}A_{m}^{\dag}A_{m}+\hbar\sum_{m}\sum
_{k}\omega_{mk}b_{mk}^{\dag}b_{mk}\text{,}\label{6a}\\
V  &  =\hbar\sum_{m,n}\sum_{k}C_{nm}V_{mk}\left(  b_{mk}^{\dag}A_{n}%
+b_{mk}A_{n}^{\dag}\right)  \text{.} \label{6b}%
\end{align}
With the diagonalized Hamiltonian $H_{0}$ we are ready to introduce the
interaction picture, defined by the transformation $U(t)=\exp\left(
-iH_{0}t/\hbar\right)  $, in which
\end{subequations}
\begin{equation}
V(t)=\hbar\sum_{m,n}\left(  \mathcal{O}_{mn}(t)A_{n}^{\dag}+\mathcal{O}%
_{mn}^{\dag}(t)A_{n}\right)  \text{,} \label{7}%
\end{equation}
where $\mathcal{O}_{mn}(t)=C_{nm}\sum_{k}V_{mk}\exp\left[  -i\left(
\omega_{mk}-\varpi_{n}\right)  t\right]  b_{mk}$. Next, we assume the
interactions between the resonators and the reservoirs to be weak enough in
order to perform a second-order perturbation approximation followed by tracing
out the reservoir degrees of freedom. We also assume a Markovian reservoir
such that the density operator of the global system can be factorized as
$\rho_{1\ldots N}(t)\otimes\rho_{R}(0)$. Under these assumptions we obtain the
reduced density operator of the network of $N$ dissipative coupled resonators
given by%
\begin{equation}
\frac{\operatorname*{d}\rho_{1,\ldots,N}(t)}{\operatorname*{d}t}=-\frac
{1}{\hbar^{2}}\int_{0}^{t}\operatorname*{d}t^{\prime}\operatorname*{Tr}%
\nolimits_{R}\left[  V(t),\left[  V(t^{\prime}),\rho_{R}(0)\otimes
\rho_{1,\ldots,N}(t)\right]  \right]  \text{.} \label{8}%
\end{equation}
Since for a thermal reservoir $\left\langle b_{mk}b_{nk^{\prime}}\right\rangle
=\left\langle b_{mk}^{\dag}b_{nk^{\prime}}^{\dag}\right\rangle =0$, we have to
solve the integrals appearing in Eq. (\ref{8}), related to correlation
functions of the form%
\begin{align}
\int_{0}^{t}\operatorname*{d}t^{\prime}\left\langle \mathcal{O}_{mn}^{\dag
}(t)\mathcal{O}_{m^{\prime}n^{\prime}}(t^{\prime})\right\rangle  &
=C_{nm}C_{n^{\prime}m^{\prime}}\int_{0}^{t}\operatorname*{d}t^{\prime}%
\sum_{k,k^{\prime}}V_{mk}V_{m^{\prime}k^{\prime}}\left\langle b_{mk}^{\dag
}b_{m^{\prime}k^{\prime}}\right\rangle \nonumber\\
&  \times\exp\left\{  i\left[  \left(  \omega_{mk}-\varpi_{n}\right)
t-\left(  \omega_{m^{\prime}k^{\prime}}-\varpi_{n^{\prime}}\right)  t^{\prime
}\right]  \right\}  \text{.} \label{9}%
\end{align}
Considering that the reservoir frequencies are very closely spaced to allow a
continuum summation and defining the average excitation of the $m$th mode
associated to the $k$th reservoir $\mathbf{N}_{m}\left(  \nu\right)  $ as
$\left\langle b_{mk}^{\dag}(\nu)b_{m^{\prime}k^{\prime}}(\nu^{\prime
})\right\rangle =2\pi\delta_{mm^{\prime}}\mathbf{N}_{m}(\nu)\delta\left(
\nu-\nu^{\prime}\right)  $, we obtain%
\begin{align}
\int_{0}^{t}\operatorname*{d}t^{\prime}\left\langle \mathcal{O}_{mn}^{\dag
}(t)\mathcal{O}_{m^{\prime}n^{\prime}}(t^{\prime})\right\rangle  &
=\delta_{mm^{\prime}}C_{nm}C_{n^{\prime}m}\operatorname*{e}\nolimits^{i\left(
\varpi_{n^{\prime}}-\varpi_{n}\right)  t}\nonumber\\
&  \times\int_{0}^{t}\operatorname*{d}t^{\prime}\int_{0}^{\infty}%
\frac{\operatorname*{d}\nu}{2\pi}\left[  V_{m}(\nu)\sigma_{m}(\nu)\right]
^{2}\mathbf{N}_{m}(\nu)\operatorname*{e}\nolimits^{-i\left(  \nu
-\varpi_{n^{\prime}}\right)  \left(  t^{\prime}-t\right)  }\text{,} \label{10}%
\end{align}
with $\sigma_{m}(\omega_{mk})$ being the density of states of the $m$th
reservoir. Assuming, as usual, that $V_{m}(\varpi_{n})$, $\sigma_{m}%
(\varpi_{n})$ and $\mathbf{N}_{m}(\varpi_{n})$ are slowly varying functions,
we obtain after the variable transformations $\varepsilon=\nu-\varpi
_{n^{\prime}}$ and $\tau=t-t^{\prime}$ the simplified form%
\begin{equation}
\int_{0}^{t}\operatorname*{d}t^{\prime}\left\langle \mathcal{O}_{mn}^{\dag
}(t)\mathcal{O}_{m^{\prime}n^{\prime}}(t^{\prime})\right\rangle =\frac{N}%
{2}\delta_{mm^{\prime}}C_{nm}C_{n^{\prime}m}\gamma_{m}(\varpi_{n^{\prime}%
})\mathbf{N}_{m}(\varpi_{n^{\prime}})\exp\left[  i\left(  \varpi_{n^{\prime}%
}-\varpi_{n}\right)  t\right]  \text{,} \label{11}%
\end{equation}
where we have defined the damping rates as%
\begin{equation}
\gamma_{m}(\varpi_{n})=\frac{1}{N}\left[  V_{m}(\varpi_{n})\sigma_{m}%
(\varpi_{n})\right]  ^{2}\int_{-R_{n}}^{\infty}\delta\left(  \varepsilon
\right)  \operatorname*{d}\varepsilon\text{.} \label{11i}%
\end{equation}

Back to the Schr\"{o}dinger picture and to the original field operators
$a_{m}$, we finally obtain from the steps outlined above, the master equation%
\begin{align}
\frac{\operatorname*{d}\rho_{1,\ldots,N}(t)}{\operatorname*{d}t}  &  =\frac
{i}{\hbar}\left[  \rho_{1,\ldots,N}(t),H_{0}\right]  +\frac{N}{2}%
\sum_{m,n,n^{\prime}}C_{n^{\prime}m}C_{nn^{\prime}}\gamma_{m}(\varpi
_{n^{\prime}})\nonumber\\
&  \times\left\{  \mathbf{N}_{m}(\varpi_{n^{\prime}})\left(  \left[
a_{n}^{\dag}\rho_{1,\ldots,N}(t),a_{m}\right]  +\left[  a_{m}^{\dag}%
,\rho_{1,\ldots,N}(t)a_{n}\right]  \right)  \right. \nonumber\\
&  \left.  +\left(  \mathbf{N}_{m}(\varpi_{n^{\prime}})+1\right)  \left(
\left[  a_{n}\rho_{1,\ldots,N}(t),a_{m}^{\dag}\right]  +\left[  a_{m}%
,\rho_{1,\ldots,N}(t)a_{n}^{\dag}\right]  \right)  \right\}  \text{.}
\label{12}%
\end{align}
\qquad

From here on we shall focus on the case of reservoirs at 0K, leaving for the
last but one section a brief analysis of the effect of finite temperatures.
Defining the effective damping matrix whose elements are%
\begin{equation}
\Gamma_{mn}=N\sum_{n^{\prime}}C_{n^{\prime}m}\gamma_{m}(\varpi_{n^{\prime}%
})C_{n^{\prime}n}\text{,} \label{13}%
\end{equation}
the master equation for the reservoirs at 0K simplifies to the generalized
Lindblad form%
\begin{align}
\frac{\operatorname*{d}\rho_{1,\ldots,N}(t)}{\operatorname*{d}t}  &  =\frac
{i}{\hbar}\left[  \rho_{1,\ldots,N}(t),H_{0}\right] \nonumber\\
&  +\sum_{m,n}\frac{\Gamma_{mn}}{2}\left(  \left[  a_{n}\rho_{1,\ldots
,N}(t),a_{m}^{\dag}\right]  +\left[  a_{m},\rho_{1,\ldots,N}(t)a_{n}^{\dag
}\right]  \right) \nonumber\\
&  \equiv\frac{i}{\hbar}\left[  \rho_{1,\ldots,N}(t),H_{0}\right]  +\sum
_{m,n}\mathcal{L}_{mn}\rho_{1,\ldots,N}(t)\mathrm{,} \label{14}%
\end{align}
where $\mathcal{L}_{mn}\rho_{1,\ldots,N}(t)$ are the Liouville operators
accounting for the direct ($m=n$) and indirect ($m\neq n$)\ dissipative
channels, respectively. Through the direct dissipative channels the
oscillators lose excitation to their own reservoirs, whereas through the
indirect channels they lose excitation to all the other reservoirs but not to
their own. We observe that for Markovian white noise reservoirs, where the
spectral densities of the reservoirs are invariant over translation in
frequency space, such that $\gamma_{m}(\varpi_{n^{\prime}})=\gamma_{m}$, the
expression (\ref{13}) reduces to $\Gamma_{mn}=N\gamma_{m}\delta_{mn}$. For
these particular reservoirs the indirect channels disappears. We also observe
that, in the weak coupling regime where $N\left\{  \lambda_{mn}\right\}
\ll\left\{  \omega_{m^{\prime}}\right\}  $ \cite{Mickel1,Mickel2}\ and
consequently $\gamma_{m}(\varpi_{n^{\prime}})\approx\gamma_{m}(\omega_{m}%
)$,\ we obtain $\Gamma_{mn}=N\gamma_{m}(\omega_{m})\delta_{mn}$, such that the
indirect channels again disappears. Therefore, it is worth nothing that the
indirect channels play a significant role only in the strong coupling regime
where $N\left\{  \lambda_{mn}\right\}  \approx\left\{  \omega_{m^{\prime}%
}\right\}  $.

\section{The Glauber-Sudarshan $P$ function}

The evolution equation for the Glauber-Sudarshan $P$ function, derived from
the master equation (\ref{14}), is given by%
\begin{equation}
\frac{\operatorname*{d}P_{1,\ldots,N}(\{\eta_{m^{\prime}}\},t)}%
{\operatorname*{d}t}=\sum_{m}\left(  \frac{\Gamma_{mm}}{2}+\sum_{n}%
\mathcal{H}_{mn}^{D}\eta_{n}\frac{\partial}{\partial\eta_{m}}+c.c.\right)
P_{1,\ldots,N}(\{\eta_{m^{\prime}}\},t)\text{,} \label{15}%
\end{equation}
where we have defined the matrix $\mathcal{H}^{D}$, with the elements%
\begin{equation}
\mathcal{H}_{mn}^{D}=\Gamma_{mn}/2+i\mathcal{H}_{mn}\text{,} \label{15i}%
\end{equation}
thus generalizing the former matrix $\mathcal{H}$ (\ref{Sym}) to account for
the dissipative ($D$) process. With the transformation $P_{1,\ldots,N}%
(\{\eta_{m^{\prime}}\},t)=\widetilde{P}_{1,\ldots,N}(\{\eta_{m^{\prime}%
}\},t)\exp\left(
{\textstyle\sum_{m}}
\Gamma_{mm}t\right)  $, and assuming a solution of Eq. (\ref{15}) of the form
$\tilde{P}(\left\{  \eta_{n}\right\}  ,t)=\tilde{P}(\left\{  \eta
_{n}(t)\right\}  )$, we obtain the differential equation%
\begin{align}
\frac{\operatorname*{d}\widetilde{P}_{1,\ldots,N}(\{\eta_{m^{\prime}}%
(t)\})}{\operatorname*{d}t}  &  =\sum_{m}\left(  \frac{\partial\eta_{m}%
(t)}{\partial t}\frac{\partial}{\partial\eta_{m}}+c.c\right)  \widetilde
{P}_{1,\ldots,N}(\{\eta_{m^{\prime}}\},t)\nonumber\\
&  =\sum_{m}\left(  \sum_{n}\mathcal{H}_{mn}^{D}\eta_{n}\frac{\partial
}{\partial\eta_{m}}+c.c.\right)  \widetilde{P}_{1,\ldots,N}(\{\eta_{m^{\prime
}}\},t)\text{,} \label{16}%
\end{align}
which makes possible to calculate the time evolution of the parameters
$\eta_{m}(t)$ through the physical quantities of the system appearing on the
elements $\mathcal{H}_{mn}^{D}$, as%
\begin{equation}
\frac{\partial\eta_{m}(t)}{\partial t}=\sum_{n}\mathcal{H}_{mn}^{D}\eta
_{n}\text{.} \label{17}%
\end{equation}

Through the transformation $\widetilde{\eta}_{m}(t)=\sum_{n}D_{mn}^{-1}%
\eta_{m}(t)$, we diagonalize the matrix $\mathcal{H}^{D}$ thus reducing Eq.
(\ref{17}) to the diagonal form $\partial\widetilde{\eta}_{m}(t)/\partial
t=\Omega_{m}\widetilde{\eta}_{m}$, whose solution is $\widetilde{\eta}%
_{m}(t)=\mathcal{A}_{m}\exp\left(  \Omega_{m}t\right)  $. Therefore, back to
the parameters $\eta_{m}(t)$ we obtain%
\begin{equation}
\eta_{m}(t)=\sum_{n}D_{mn}\widetilde{\eta}_{n}(t)=\sum_{n}D_{mn}\exp\left(
\Omega_{n}t\right)  \mathcal{A}_{n}\text{,} \label{18}%
\end{equation}
where the elements of the $m$th column of matrix $\mathbf{D}$ define the $m$th
eigenvector associated to the eigenvalue $\Omega_{m}$ of matrix $\mathcal{H}%
^{D}$, and by setting the initial condition $\eta_{m}(t=0)\equiv\eta_{m}^{0}$,
we verify from Eq. (\ref{18}) that $\sum_{n}D_{mn}\mathcal{A}_{n}=\eta_{m}%
^{0}$. Therefore, $\mathcal{A}_{n}=\sum_{m}D_{nm}^{-1}\eta_{m}^{0}$ and,
consequently%
\begin{equation}
\eta_{m}(t)=\sum_{m^{\prime},n}D_{mn}\exp\left(  \Omega_{n}t\right)
D_{nm^{\prime}}^{-1}\eta_{m^{\prime}}^{0}\text{,} \label{19}%
\end{equation}
leading to the solution for the Glauber-Sudarshan $P$-function%
\begin{equation}
P_{1,\ldots,N}(\{\eta_{n}\},t)=\exp\left(  \sum_{m}\Gamma_{mm}t\right)
\left.  P_{1,\ldots,N}(\{\eta_{n}\},t=0)\right\vert _{\left\{  \eta
_{n}\right\}  \rightarrow\left\{  \eta_{n}(t)\right\}  }\text{.} \label{20}%
\end{equation}
Therefore, having the $P$-function at time $t=0$, we immediately obtain it at
any other time by substituting the set $\left\{  \eta_{n}\right\}  $ by
$\left\{  \eta_{n}(t)\right\}  $.

\subsection{From the general matrix $\mathcal{H}^{D}$ to particular
dissipative topologies}

In subsection II.B we illustrate how to construct particular nondissipative
topologies from a general symmetric network described by matrix (\ref{Sym}).
Now, after introducing the generalized matrix $\mathcal{H}^{D}$ we are in the
position to enlarge the focus by constructing networks entirely composed of
dissipative oscillators or, in a more general fashion, composed by mixed
nondissipative and dissipative oscillators. Back to the symmetric, central,
circular and linear networks, when considering that they are all composed by
dissipative oscillators, each one coupled to its respective reservoir, we
obtain for the matrices $\mathcal{H}^{D}$ exactly the same structure as those
in subsection 2.B. However, from Eq. (\ref{15i}) we verify that the matrix
elements $\mathcal{H}_{mn}^{D}$ follow from those of $\mathcal{H}_{mn}$
multiplied by the imaginary $i$ apart from the correction $\Gamma_{mn}/2$
coming from the dissipative process. As an example, for the symmetric network
composed entirely by dissipative oscillators, as sketched in Fig. 2 (a), the
Hamiltonian $\mathcal{H}^{D}$ assumes the form%
\begin{equation}
\mathcal{H}_{sym}^{D}=i\mathcal{H}_{sym}+\frac{1}{2}\left(
\begin{array}
[c]{ccccc}%
\Gamma_{11} & \Gamma_{12} & \Gamma_{13} & \cdots & \Gamma_{1N}\\
\Gamma_{21} & \Gamma_{22} & \Gamma_{23} & \cdots & \Gamma_{2N}\\
\Gamma_{31} & \Gamma_{32} & \Gamma_{33} & \cdots & \Gamma_{3N}\\
\vdots & \vdots & \vdots & \ddots & \vdots\\
\Gamma_{N1} & \Gamma_{N2} & \Gamma_{N3} & \cdots & \Gamma_{NN}%
\end{array}
\right)  \text{.} \label{SymD}%
\end{equation}
Let us consider, instead, the case of a mixed symmetric ($mix-sym$) network,
composed by an even total number of oscillators $N$, where those designated by
odd (even) numbers are nondissipative (dissipative), as sketched in Fig. 2
(b). In this case, the Hamiltonian $\mathcal{H}^{D}$ is given by%
\begin{equation}
\mathcal{H}_{mix-sym}^{D}=i\mathcal{H}_{sym}+\frac{1}{2}\left(
\begin{array}
[c]{ccccc}%
0 & 0 & 0 & \cdots & 0\\
\Gamma_{21} & \Gamma_{22} & \Gamma_{23} & \cdots & \Gamma_{2N}\\
0 & 0 & 0 & \cdots & 0\\
\vdots & \vdots & \vdots & \ddots & \vdots\\
\Gamma_{N1} & \Gamma_{N2} & \Gamma_{N3} & \cdots & \Gamma_{NN}%
\end{array}
\right)  \text{.} \label{MixSymD}%
\end{equation}

As a last example of a mixed network we consider the topology sketched in Fig.
2 (c), where three linear chains of coupled oscillators are connected
together, with the central (lateral) chain(s) being composed by nondissipative
(dissipative) oscillators. In this case, labeling the oscillators as in Fig. 2
(c), we obtain%
\begin{equation}
\mathcal{H}_{mix}^{D}=i\left(
\begin{array}
[c]{ccccccccc}%
\omega_{1} & \lambda_{12} & 0 & \lambda_{14} &  &  &  &  & \\
\lambda_{12} & \omega_{2} & \lambda_{23} & 0 & \lambda_{25} &  &  &
\textbf{\Huge{0}}%
& \\
0 & \lambda_{23} & \omega_{3} & 0 & 0 & \lambda_{36} &  &  & \\
\lambda_{14} & 0 & 0 & \omega_{4} & \lambda_{45} & 0 & \lambda_{47} &  & \\
& \lambda_{25} & 0 & \lambda_{45} & \omega_{5} & \lambda_{56} & 0 &
\lambda_{58} & \\
&  & \lambda_{36} & 0 & \lambda_{56} & \omega_{6} & 0 & 0 & \lambda_{69}\\
&  &  & \lambda_{47} & 0 & 0 & \omega_{7} & \lambda_{78} & 0\\
&
\textbf{\Huge{0}}%
&  &  & \lambda_{58} & 0 & \lambda_{78} & \omega_{8} & \lambda_{89}\\
&  &  &  &  & \lambda_{69} & 0 & \lambda_{89} & \omega_{9}%
\end{array}
\right)  +\frac{1}{2}\left(
\begin{array}
[c]{ccccc}%
\Gamma_{11} & \Gamma_{12} & \Gamma_{13} & \cdots & \Gamma_{19}\\
\Gamma_{21} & \Gamma_{22} & \Gamma_{23} & \cdots & \Gamma_{29}\\
0 & 0 & 0 & \cdots & 0\\
\Gamma_{41} & \Gamma_{42} & \Gamma_{43} & \cdots & \Gamma_{49}\\
0 & 0 & 0 & \cdots & 0\\
\Gamma_{61} & \Gamma_{62} & \Gamma_{63} & \cdots & \Gamma_{69}\\
0 & 0 & 0 & \cdots & 0\\
\Gamma_{81} & \Gamma_{82} & \Gamma_{83} & \cdots & \Gamma_{89}\\
\Gamma_{91} & \Gamma_{92} & \Gamma_{93} & \cdots & \Gamma_{99}%
\end{array}
\right)  \text{.} \label{Mix}%
\end{equation}

Therefore, for a given topology, the diagonalization of the matrix
$\mathcal{H}^{D}$, leading to the eigenvalues $\Omega_{m}$ and eigenvectors
composing the matrix $\mathbf{D}$, enables us to obtain the evolved
Glauber-Sudarshan $P$-function and, consequently, the reduced density operator
of the network, given by%
\begin{equation}
\rho_{1,\ldots,N}(t)=\left[  \bigotimes\limits_{m}\int d^{2}\eta_{m}%
^{0}\left\vert \eta_{m}^{0}\right\rangle \left\langle \eta_{m}^{0}\right\vert
\right]  P_{1,\ldots,N}(\{\eta_{n}\},t)\text{.} \label{Rho}%
\end{equation}

\section{Initial states, decoherence, and linear entropy}

Starting from two different initial states of the network, given by general
superpositions of coherent and Fock states, in this section we obtain the
evolved reduced density operator (\ref{Rho}) for both cases.

\subsection{A general superposition of coherent states}

Let us assume that the initial pure state of the network is given by%
\begin{equation}
\rho(0)=\mathcal{N}^{2}\sum_{r,s=1}^{Q}\Lambda_{r}\Lambda_{s}^{\ast}\left\vert
\left\{  \beta_{m}^{r}\right\}  \right\rangle \left\langle \left\{  \beta
_{m}^{s}\right\}  \right\vert \text{,} \label{21}%
\end{equation}
where $\mathcal{N}$ is the normalization factor, $\Lambda_{r}$ is the
probability amplitudes of the product state $\left\vert \left\{  \beta_{m}%
^{r}\right\}  \right\rangle =\bigotimes\nolimits_{m=1}^{N}\left\vert \beta
_{m}^{r}\right\rangle $, and the labels $r$ and $s$ run from $1$\ to the
integer $Q$. The superscript $r$ stands for the $r$th state of the
superposition while the subscript $m$ stands for the coherent state of $m$th
oscillator. We stress that the discrete sum of product states in Eq.
(\ref{21}) can be substituted by the continuum sum $\left\vert \psi
(0)\right\rangle =\mathcal{N}\int d\theta\Lambda\left(  \theta\right)
\left\vert \left\{  \beta_{m}\left(  \theta\right)  \right\}  \right\rangle $
with no further complication. After some algebra, we verify that the state
(\ref{21}) evolves to the $N$-oscillator density operator%
\begin{equation}
\rho_{1,\ldots,N}(t)=\mathcal{N}^{2}\sum_{r,s}\Lambda_{r}\Lambda_{s}^{\ast
}\frac{\left\langle \left\{  \beta_{m}^{s}\right\}  \right.  \left\vert
\left\{  \beta_{m}^{r}\right\}  \right\rangle }{\left\langle \left\{
\zeta_{m}^{s}(t)\right\}  \left\vert \left\{  \zeta_{m}^{r}(t)\right\}
\right.  \right\rangle }\left\vert \left\{  \zeta_{m}^{r}(t)\right\}
\right\rangle \left\langle \left\{  \zeta_{m}^{s}(t)\right\}  \right\vert
\text{,} \label{22}%
\end{equation}
where the excitation of the $m$th oscillator\ is given by%
\begin{equation}
\zeta_{m}^{r}\left(  t\right)  =\sum_{n}\Theta_{mn}(t)\beta_{n}^{r} \label{23}%
\end{equation}
with the time-dependent matrix elements%
\begin{equation}
\Theta_{mn}(t)=\sum_{m^{\prime}}D_{mm^{\prime}}\exp\left(  -\Omega_{m^{\prime
}}t\right)  D_{m^{\prime}n}^{-1}\text{.} \label{24}%
\end{equation}
For the reduced density operator of the $m$th oscillator we obtain%
\begin{equation}
\rho_{m}(t)=\mathcal{N}^{2}\sum_{r,s}\Lambda_{r}\Lambda_{s}^{\ast}%
\frac{\left\langle \left\{  \beta_{n}^{s}\right\}  \right.  \left\vert
\left\{  \beta_{n}^{r}\right\}  \right\rangle }{\left\langle \zeta_{m}%
^{s}(t)\left\vert \zeta_{m}^{r}(t)\right.  \right\rangle }\left\vert \zeta
_{m}^{r}(t)\right\rangle \left\langle \zeta_{m}^{s}(t)\right\vert \text{,}
\label{25}%
\end{equation}
where the influence of all the other oscillators of the network is present
explicitly in the product $\left\langle \left\{  \beta_{n}^{s}\right\}
\right.  \left\vert \left\{  \beta_{n}^{r}\right\}  \right\rangle $ and
implicitly in the states $\left\vert \zeta_{m}^{r}(t)\right\rangle $.

\subsection{A general superposition of Fock states}

In the Fock basis we consider the state%
\begin{equation}
\left\vert \varphi(0)\right\rangle =%
{\displaystyle\sum\limits_{n_{1},\ldots,n_{N}}}
C_{n_{1},\ldots,n_{N}}\left\vert n_{1},\ldots,n_{N}\right\rangle \text{,}
\label{26}%
\end{equation}
where $n_{m}$ stands for the photon number of $m$th oscillator and
$C_{n_{1},\ldots,n_{N}}$ is the probability amplitude of each state in the
superposition. After a lengthy calculation we verify that the $N$-oscillator
density operator is given by%
\begin{align}
\rho_{1,\ldots,N}(t)  &  =%
{\displaystyle\sum\limits_{n_{1},\ldots,n_{N}}}
{\displaystyle\sum\limits_{m_{1},\ldots,m_{N}}}
C_{m_{1},\ldots,m_{N}}^{\ast}C_{n_{1},\ldots,n_{N}}\left(
{\displaystyle\prod\limits_{m}}
\sum_{q_{m}=0}^{n_{m}}\sum_{k_{m}=0}^{\infty}\frac{\left(  -1\right)  ^{k_{m}%
}\sqrt{m_{m}!n_{m}!}}{\left(  n_{m}-q_{m}\right)  !k_{m}!}\right) \nonumber\\
&  \times\left\vert \mathcal{F}(\left\{  q_{\ell}\right\}  ,\left\{  k_{\ell
}\right\}  ,t)\right\rangle \left\langle \mathcal{F}(\left\{  m_{\ell}%
-n_{\ell}+q_{\ell}\right\}  ,\left\{  k_{\ell}\right\}  ,t)\right\vert
\label{26.5}%
\end{align}
where%
\begin{align}
\left\vert \mathcal{F}(\left\{  q_{\ell}\right\}  ,\left\{  k_{\ell}\right\}
,t)\right\rangle  &  =\bigotimes\limits_{m}\sum_{j_{m}=0}^{\infty}%
\frac{\left(  j_{m}+k_{m}\right)  !}{\sqrt{j_{m}!}}\sum_{\mu_{m,1}=0}%
^{j_{m}+k_{m}}\frac{\left(  \Theta_{m,1}(t)\right)  ^{j_{m}+k_{m}-\mu_{m,1}%
}\left(  \Theta_{m,N}(t)\right)  ^{\mu_{m,N-1}}}{\left(  j_{m}+k_{m}-\mu
_{m,1}\right)  !\mu_{m,N-1}!}\nonumber\\
&  \times\left(
{\displaystyle\prod\limits_{i=2}^{N-1}}
\sum_{\mu_{m,i}=0}^{\mu_{m,i-1}}\frac{\left(  \Theta_{m,i}(t)\right)
^{\mu_{m,i-1}-\mu_{m,i}}}{\left(  \mu_{m,i-1}-\mu_{m,i}\right)  !}\right)
\nonumber\\
&  \times\delta\left\{  \sum_{n}\left[  j_{n}+k_{n}-\mu_{n,m}\left(
1-\delta_{m,N}\right)  \right]  -\sum_{r=1}^{m}q_{r}\right\}  \left\vert
j_{m}\right\rangle \label{27}%
\end{align}
where $\delta\left(  x\right)  $ equals unity for $x=0$, being null otherwise.

\subsection{State transfer and recurrence dynamics}

From the reduced density operators $\rho_{m}(t)$ following from Eqs.
(\ref{22}) and (\ref{26.5}) it is directly to verify the transfer of an
initial state prepared in the $m$th oscillator to the remaining one of the
network, followed by the recurrence of this state back to the $m$th
oscillator. The probability of recurrence of an initial state $\rho_{m}(0)$
prepared in the $m$th oscillator is given by the expression
\begin{equation}
\mathcal{P}_{R}(t)\equiv Tr\left[  \rho_{m}(t)\rho_{m}(0)\right]  \text{,}
\label{i}%
\end{equation}
which is also a measurement of the fidelity of the initial state $\rho_{m}%
(0)$, expected to decrease due to the dissipative process. For the probability
of transfer of the initial state $\rho_{m}(0)$ to a particular $n$th
oscillator picked up from the remaining $N-1$ of the network, we get%
\begin{equation}
\mathcal{P}_{T}(t)\equiv Tr\left[  \rho_{n}(t)\rho_{m}(0)\right]  \text{.}
\label{ii}%
\end{equation}

From Eqs. (\ref{i}) and (\ref{ii}) it can be verified --- as analyzed in
details in Refs. \cite{Mickel1} and \cite{Mickel2} for the particular
symmetric and central topologies, respectively --- that an initial
superposition prepared in the $m$th oscillator bounces between its original
oscillators and all the remaining oscillators of the network. Evidently, the
dynamics of a given prepared state through the network can be manipulated
through the choice of the topology.

\subsection{Decoherence}

From a given initial superposition of coherent states (\ref{21}) and the
density operator of the network (\ref{22}), we can estimate the decoherence
time of an arbitrary chosen off-diagonal element of the density operator
relatively to the relaxation time of the diagonal elements. In fact, the
literature concerned with the decoherence of $N$-dimensional superpositions
acquaint only for relative time-decay measurements by which the larger the
distance from the main diagonal of the matrix elements, the smaller are their
decay time \cite{MMC}. From Eq. (\ref{22}) we verify that such relative
time-decay measurements follows from the real part of the coefficients
$\left\langle \left\{  \beta_{m}^{r}\right\}  \right.  \left\vert \left\{
\beta_{m}^{s}\right\}  \right\rangle /\left\langle \left\{  \zeta_{m}%
^{r}\left(  t\right)  \right\}  \left\vert \left\{  \zeta_{m}^{s}\left(
t\right)  \right\}  \right.  \right\rangle $ which is directly computed from
the initial state of the network together with Eq. (\ref{23}).

However, additional ingredients concerning the decoherence dynamics arise when
considering a network of dissipative quantum systems. In Ref. \cite{Mickel},
where a minimal network of two dissipative oscillators is analyzed, it is
demonstrated that the decoherence time of a "Schr\"{o}dinger cat"-like state
prepared in one of the oscillators can be doubled compared to that when the
same state is prepared in an isolated dissipative oscillator. This result
follows when the decay rate of the oscillator, where the state is prepared, is
significantly larger than the other one composing the network. A generalized
analysis of decoherence for the case of a symmetric network of dissipative
oscillators is presented in Ref. \cite{Mickel3}, where the physical
ingredients that enable the emergence of relaxation-free and decoherence-free
subspaces are exposed. On this regard, a detailed study of the optimum
topologies leading to maximum decoherence times of superposition states
prepared in particular oscillators of dissipative networks will be presented
elsewhere \cite{Mickel4}. The memory devices presented in Ref. \cite{Mickel4},
which follows from the general formalism presented here, combines both
ingredients: $i)$ the large decay rate of the storage oscillators of the
network ---\ those except the one where the state to be protected is prepared
--- and $ii)$ specific dynamics of this state through the network, achieved by
engineering particular topologies.

\subsection{Linear entropy}

From the density operator in Eq. (\ref{22}), we are able to calculate the
linear entropies for the mixed states of the whole network, $\mathcal{S}%
_{1,\ldots,N}(t)$, of oscillator $1$ (or any other particular oscillator),
$\mathcal{S}_{1}(t)$, and of all the remaining $N-1$ oscillators,
$\mathcal{S}_{2,\ldots,N}(t)$. which are given by
\begin{subequations}
\label{E1}%
\begin{align}
\mathcal{S}_{1,\ldots,N}(t)  &  =1-\mathrm{Tr}_{1,\ldots,N}\left[
\rho_{1,\ldots,N}(t)\right]  ^{2}\nonumber\\
&  =1-\mathcal{N}^{4}\sum_{r,s,p,q}\left\langle \left\{  \beta_{m}%
^{r}\right\}  \right.  \left\vert \left\{  \beta_{m}^{s}\right\}
\right\rangle \left\langle \left\{  \beta_{m}^{p}\right\}  \right.  \left\vert
\left\{  \beta_{m}^{q}\right\}  \right\rangle \nonumber\\
&  \times\exp\left[  -\sum_{n}\left(  \beta_{n}^{s}-\beta_{n}^{q}\right)
\left(  \beta_{n}^{r}-\beta_{n}^{p}\right)  ^{\ast}\sum_{m}\left\vert
\Theta_{mn}(-t)\right\vert ^{2}\right]  \text{,}\label{E1a}\\
\mathcal{S}_{1}(t)  &  =1-\mathrm{Tr}_{1}\left[  \rho_{1}(t)\right]
^{2}\nonumber\\
&  =1-\mathcal{N}^{4}\sum_{r,s,p,q}\left\langle \left\{  \beta_{m}%
^{r}\right\}  \right.  \left\vert \left\{  \beta_{m}^{s}\right\}
\right\rangle \left\langle \left\{  \beta_{m}^{p}\right\}  \right.  \left\vert
\left\{  \beta_{m}^{q}\right\}  \right\rangle \nonumber\\
&  \times\exp\left[  -\sum_{n}\left(  \beta_{n}^{s}-\beta_{n}^{q}\right)
\left(  \beta_{n}^{r}-\beta_{n}^{p}\right)  ^{\ast}\left\vert \Theta
_{1n}(-t)\right\vert ^{2}\right]  \text{,}\label{E1b}\\
\mathcal{S}_{2,\ldots,N}(t)  &  =1-\mathrm{Tr}_{2,\ldots,N}\left[
\rho_{2,\ldots,N}(t)\right]  ^{2}\nonumber\\
&  =1-\mathcal{N}^{4}\sum_{r,s,p,q}\left\langle \left\{  \beta_{m}%
^{r}\right\}  \right.  \left\vert \left\{  \beta_{m}^{s}\right\}
\right\rangle \left\langle \left\{  \beta_{m}^{p}\right\}  \right.  \left\vert
\left\{  \beta_{m}^{q}\right\}  \right\rangle \nonumber\\
&  \times\exp\left[  -\sum_{n}\left(  \beta_{n}^{s}-\beta_{n}^{q}\right)
\left(  \beta_{n}^{r}-\beta_{n}^{p}\right)  ^{\ast}\sum_{\ell=2}^{N}\left\vert
\Theta_{\ell n}(-t)\right\vert ^{2}\right]  \text{,} \label{E1c}%
\end{align}
where the second equality follows from the initial state (\ref{21}) and, as
like as $r$ and $s$, the labels $p$ and $q$ run from $1$ to the integer $Q$.
With these expressions we can analyze, as discussed in Refs.
\cite{Mickel1,Mickel2,Mickel3}, the evolution of the correlation between the
reduced state of oscillator $1$ and that of all the remaining $N-1$
oscillators, through the excess entropy, defined as
\end{subequations}
\begin{equation}
\mathcal{E}(t)\equiv\mathcal{S}_{1}(t)+\mathcal{S}_{2,\ldots,N}(t)-\mathcal{S}%
_{1,\ldots,N}(t)\mathrm{.} \label{E2}%
\end{equation}

The excess entropy can also reveals (through a residual value of $E(t)$) the
development of an inevitable background correlation between all the network
oscillators which thus become permanently entangled
\cite{Mickel1,Mickel2,Mickel3}. This background correlation arises from two
different mechanisms: First, the action of the indirect channels ($L_{mn}%
\rho_{1,\ldots,N}(t)$) and, secondly, the action of the direct channels
($L_{mm}\rho_{1,\ldots,N}(t)$), when the decay rates $\Gamma_{mn}$ are
different from each other. For equal decay rates, the direct decay channels do
not contribute to the development of the background correlation. The indirect
channels thus play an important role in the entanglement process in
dissipative networks.

\section{The case of a common reservoir for the whole network}

In this section we extend our analysis to contemplate the case where all the
oscillators of the network are coupled to one common reservoir (at 0K). For a
particular symmetric network, the different results following from both cases
of distinct or a common reservoir have been discussed in Ref. \cite{Mickel3}
in connection with the emergence of relaxation-free and/or decoherence-free
subspaces. In Refs. \cite{Mickel,Mickel1,Mickel2,Mickel3} a brief discussion
is also provided about the rather unusual scenario of a common reservoir.

The Hamiltonian for the case of a common reservoir is given by%
\begin{align}
H  &  =\hbar\sum_{m}\omega_{m}a_{m}^{\dag}a_{m}+\frac{\hbar}{2}\sum_{m\neq
n}\lambda_{mn}\left(  a_{m}^{\dag}a_{n}+a_{m}a_{n}^{\dag}\right) \nonumber\\
&  +\sum_{k}\omega_{k}b_{k}^{\dag}b_{k}+\hbar\sum_{m}\sum_{k}V_{mk}\left(
b_{k}^{\dag}a_{m}+b_{k}a_{m}^{\dag}\right)  \text{.} \label{28}%
\end{align}
Following the same steps as in Section II, we obtain from Hamiltonian
(\ref{28}) the same master equation (\ref{14}) derived previously for the case
of distinct reservoirs, but with the effective damping matrix (\ref{13})
replaced by%
\begin{equation}
\Gamma_{mn}=N\sum_{m^{\prime},n^{\prime}}\xi_{mm^{\prime}}(\varpi_{n^{\prime}%
})C_{n^{\prime}m^{\prime}}C_{n^{\prime}n}\text{,} \label{29}%
\end{equation}
where%
\begin{equation}
\xi_{mn}(\varpi_{m^{\prime}})=\int_{0}^{t}\operatorname*{d}\tau\int
_{0}^{\infty}\frac{\operatorname*{d}\nu}{\pi}\sigma^{2}(\nu)V_{m}(\nu
)V_{n}(\nu)\operatorname*{e}\nolimits^{-i\left(  \nu-\varpi_{m^{\prime}%
}\right)  \left(  t-\tau\right)  }\text{.} \label{30}%
\end{equation}
The correlation factor $\xi_{mn}$ arises from the fact that both network
oscillators, $m$ and $n$, may interact indirectly through their common
reservoir. To analyze more closely such a correlation, we assume (as usual for
the case of weak coupling between the system and the reservoir), that the
network oscillators only interact with the reservoir modes in the neighborhood
of their normal modes. Under this assumption the maximum correlation takes
place when both oscillators $m$ and $n$ are identically coupled with the same
group of reservoir modes, i.e., when $V_{m}(\nu)=V_{n}(\nu)$.\ Otherwise, a
partial correlation arises when the coupling between the oscillators with the
reservoir modes turns out not to be identical, i.e., $\left\{  V_{m}%
(\nu)\right\}  \cap\left\{  V_{n}(\nu)\right\}  \neq$ $\varnothing$. In this
case, the oscillators may still be coupled with the same group of reservoir
modes, but with different strengths, or be coupled with different groups of
reservoir modes apart from a common intersection of them. The correlation
between the oscillators disappears only when $\left\{  V_{m}(\nu)\right\}
\cap\left\{  V_{n}(\nu)\right\}  =$ $\varnothing$, i.e., there is practically
no intersection of common reservoir modes coupled to both oscillators.

It is particularly interesting to note that in the case where $\left\{
V_{m}(\nu)\right\}  \cap\left\{  V_{n}(\nu)\right\}  =$ $\varnothing$ and
consequently $\xi_{mn}=0$, only the self-correlation $\xi_{mm}$ survives,
which reduces to the damping factor $\delta_{mn}\gamma_{m}$ in Eq.
(\ref{11i}), apart from the unique frequency distribution $\sigma(\nu)$ of the
common reservoir. By its turn, when assuming the coupling strengths $V_{m}$
between the oscillators and the common reservoir to be all different to
compensate the unique $\sigma(\nu)$, the effective damping factor (\ref{29})
arising from the self-correlations reduces to that of the case of distinct
reservoirs in Eq. (\ref{13}). Therefore, it is possible to derive the master
equation for the case of distinct reservoir starting from that of a common
one, under the condition that no correlation between two oscillators is
induced by their common reservoir. Conversely, it is also possible to shift
from the case of distinct reservoir to that of a common one assuming that all
the reservoirs presents the same frequency distribution $\sigma(\nu)$ and the
limit of strong interactions between the network oscillators. In this limit,
as discussed in Refs. \cite{Mickel,Mickel1,Mickel2,Mickel3}, the condition
$N\lambda_{mn}\gtrsim\omega_{m^{\prime}}$ must be satisfied. The interesting
aspect of such a condition is that it can be fulfilled for coupling strengths
$\lambda_{mn}\ll\omega_{m^{\prime}}$, as long as a sufficiently large network
is provided ($N\gtrsim\omega_{m^{\prime}}/\lambda_{mn}$).

\section{The case of reservoirs at finite temperatures}

Since a formal approach for the case where the reservoirs are at finite
temperatures is somewhat demanding, here we shall present a brief qualitative
analysis of this case. Our analysis focus on the normal-mode oscillators
$\varpi_{m}$ (represented by the operators $A_{m}$ and $A_{m}^{\dagger}$)
under the assumption that all the coupling strengths between the original
oscillators $\omega_{m}$ (represented by $a_{m}$ and $a_{m}^{\dagger}$) and
their respective reservoirs are around the same. We first note that the $N$
normal-mode oscillators are decoupled from each other, whereas each one
interacts with all the $N$ reservoirs as described by Hamiltonian in Eq.
(\ref{6}). Evidently, when the couplings $\lambda_{mn}$ between the original
oscillators are all turned off, the normal-mode oscillators degenerate into
the original one. Moreover, when the coupling strengths are significantly
smaller than the natural frequencies of the original oscillators, i.e.,
$N\left\{  \lambda_{mn}\right\}  \ll\left\{  \omega_{m^{\prime}}\right\}  $,
the magnitude of the interaction between the $m$th normal-mode oscillator with
the $m$th reservoir is significantly larger than those with the remaining
$N-1$ reservoirs. In fact, when $N\left\{  \lambda_{mn}\right\}  \ll\left\{
\omega_{m^{\prime}}\right\}  $, the indirect channels are not quite effective
as pointed out above. Otherwise, when $N\left\{  \lambda_{mn}\right\}
\approx\left\{  \omega_{m^{\prime}}\right\}  $, the indirect channels become
as effective as the direct one, and the magnitude of the interactions between
the $m$th normal-mode oscillator with all the reservoirs becomes quite the
same. From the above qualitative observations we next discuss the effects of
temperature in both cases of distinct reservoirs and a common reservoir.

\subsection{Distinct reservoirs}

For the case of $N$ distinct reservoirs at finite temperatures $T_{m}$, in the
regime where $N\left\{  \lambda_{mn}\right\}  \ll\left\{  \omega_{m^{\prime}%
}\right\}  $, we thus conclude that the (non null) mean energy $\left\langle
E_{m}\right\rangle $ of the $m$th normal-mode oscillator will practically be
defined by its associated $m$th reservoir. Consequently, in the steady state
configuration, each of the normal-mode oscillators presents a different mean
energy which is defined by the equilibrium reached with all the reservoirs,
but mostly with its associated reservoir. As expected, in the regime where
$N\left\{  \lambda_{mn}\right\}  \approx\left\{  \omega_{m^{\prime}}\right\}
$, all the normal-mode oscillators present approximately the same mean energy,
since the magnitudes of their couplings with the different reservoirs are
approximately the same. Summarizing, in spite of the different temperatures of
the reservoirs, in the steady state configuration of the regime where
$N\left\{  \lambda_{mn}\right\}  \approx\left\{  \omega_{m^{\prime}}\right\}
$, all the normal-mode oscillators vibrate with approximately the same mean
energy, whereas in the regime $N\left\{  \lambda_{mn}\right\}  \ll\left\{
\omega_{m^{\prime}}\right\}  $, each normal-mode oscillator exhibits a
different mean energy.

\subsection{\textbf{A common reservoir}}

When a common reservoir is considered, the scenario is much like that arising
in the case of distinct reservoirs in the regime $N\left\{  \lambda
_{mn}\right\}  \approx\left\{  \omega_{m^{\prime}}\right\}  $. In fact, as the
coupling strengths between the normal-mode oscillators and their common
reservoir are around the same, in the steady state configuration all the
normal-mode oscillators exhibt around the same mean energy.

\section{Concluding remarks}

Motivated by the necessity to better understanding the coherence and
decoherence dynamics of quantum states in networks composed by a large number
of dissipative quantum systems, we have studied chains of dissipative harmonic
oscillators. We presented previous results related to different topologies, by
starting with the simplest case of only two coupled oscillators \cite{Mickel}
and generalizing the analysis to the case of $N$ coupled oscillators in a
symmetric \cite{Mickel1} and a central-oscillator network \cite{Mickel2}. The
developments in Refs. \cite{Mickel1,Mickel2} were extended to the analysis of
the physical ingredients responsible for the emergence of decoherence-free
subspace \cite{Mickel3}.

Since in Refs. \cite{Mickel1,Mickel2} we have treated two particular
topologies independently, in the present contribution we have presented a
general formalism to treat whichever the topology of a chain of dissipative
harmonic oscillators. Starting from a symmetric network, were all the
oscillators are coupled together, apart from being coupled to their respective
reservoirs (or to a common one), we have derived the master equation and the
associated evolution equation of the Glauber-Sudarshan $P$-function. We thus
have showed how to particularize such results for whichever the specific
network. We also shown how to obtain the master equation for the case where
each oscillator is coupled to its respective reservoir starting from that
where all the oscillators are coupled to a common reservoir.

The presented formalism is quite general and can be used to compute the
decoherence time of pure or mixed states prepared in a particular oscillator
of the network or even in a cluster of oscillators of the network. The
correlation between the states of the network can also be computed through the
excess entropy defined for a bipartite system \cite{Mickel, Mickel1,Mickel2}.

It is worth stressing that an exact model to treat a dissipative quantum
oscillator can be extracted from the present treatment. To this end, we have
to pick up a single oscillator of the network --- our system of interest ---
and couple it to all the other oscillators which play the role of the
reservoir. Thus, disconsidering the reservoirs which we have treated here
through the standard perturbative approach, we end up with an exact formalism
to account for the dissipative effects over a harmonic oscillator. Evidently,
such an exact formalism can also be pursued for dissipative systems other than
a harmonic oscillator. An extensive analysis of the effects coming from the
exact treatment of dissipation emerging from this work will be presented elsewhere.

\textbf{Acknowledgments}

We wish to express thanks for the support from the Brazilian agencies FAPESP
and CNPq.

\textbf{Figure captions}

Fig. 1 Sketches of nondissipative symmetric (a), central (b), circular (c),
and linear (d) networks.

Fig. 2 Sketches of a dissipative symmetric network (a), a mixed-symmetric
network (b), composed by dissipative and nondissipative oscillators, and a
mixed network composed of dissipative and nondissipative chains of oscillators (c).

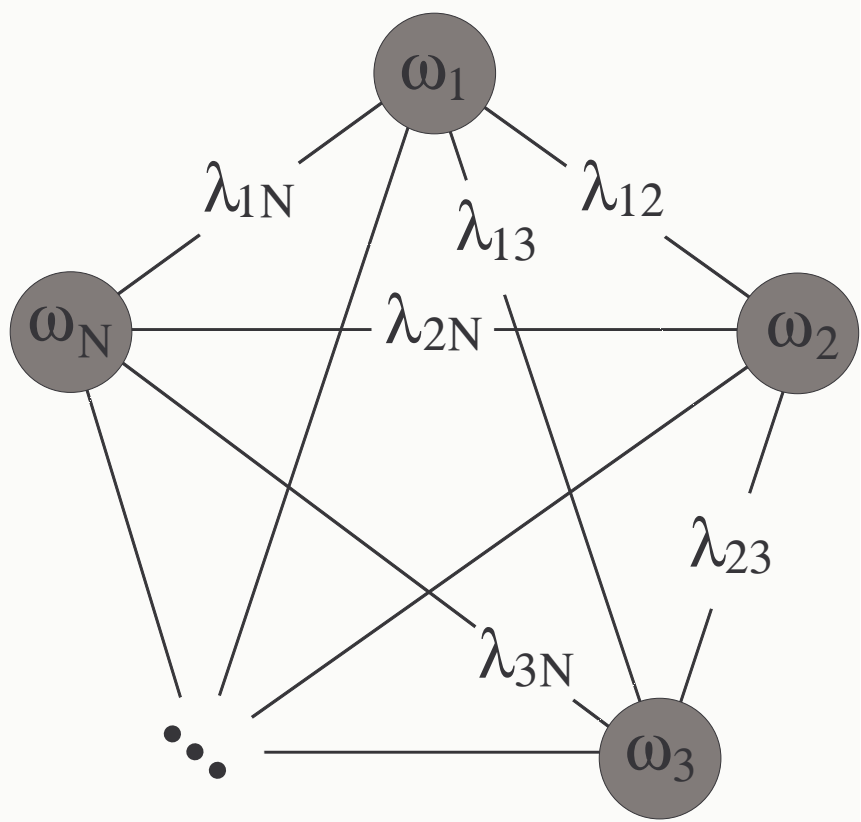
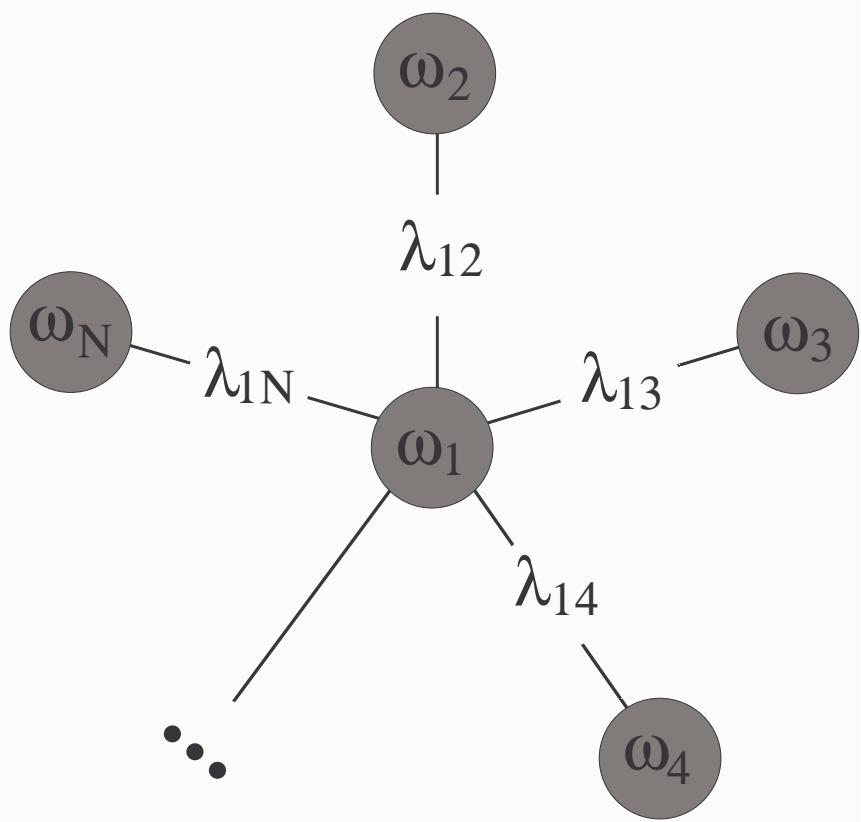
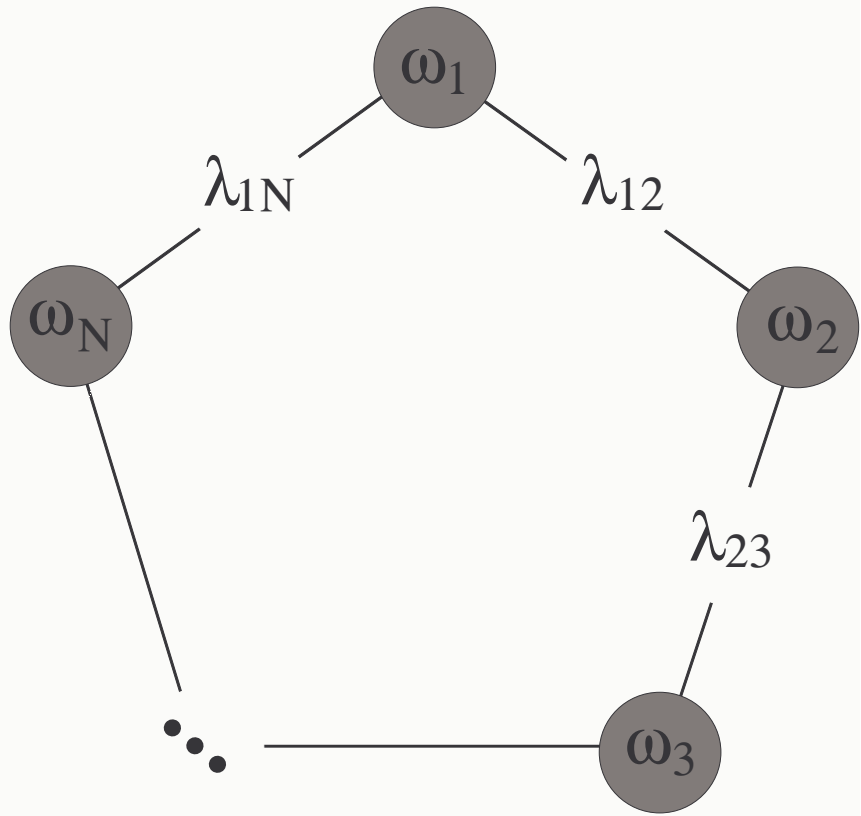
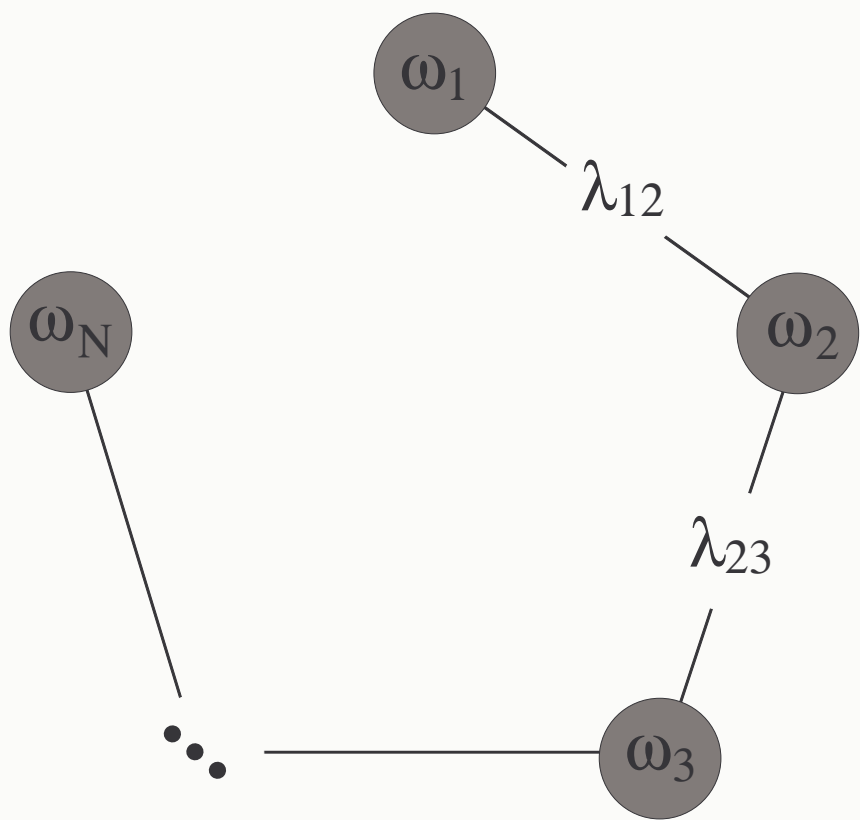
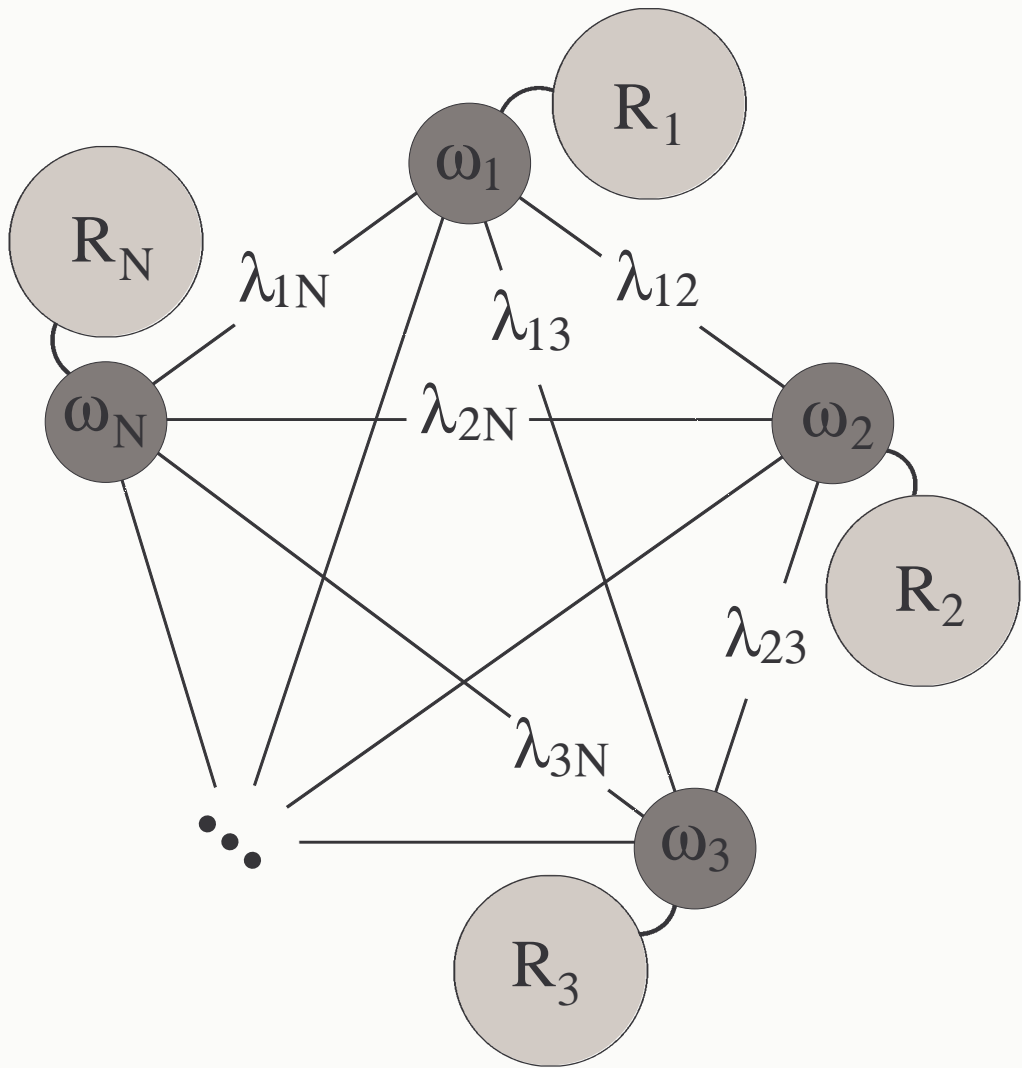
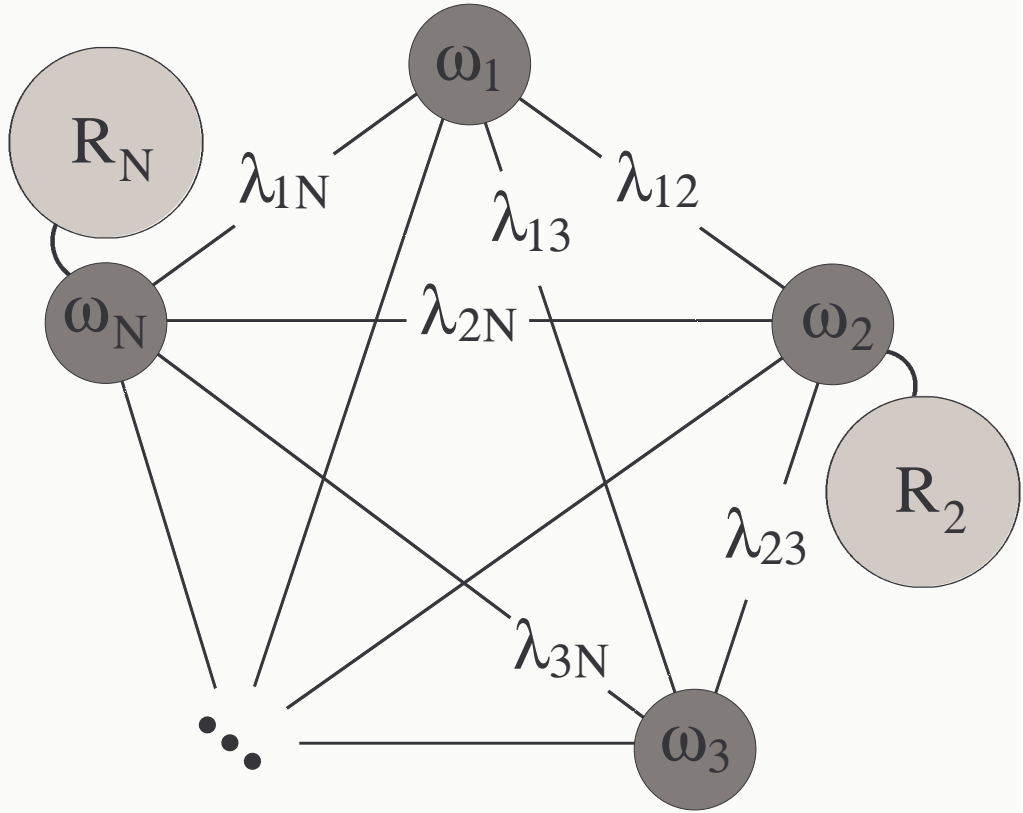
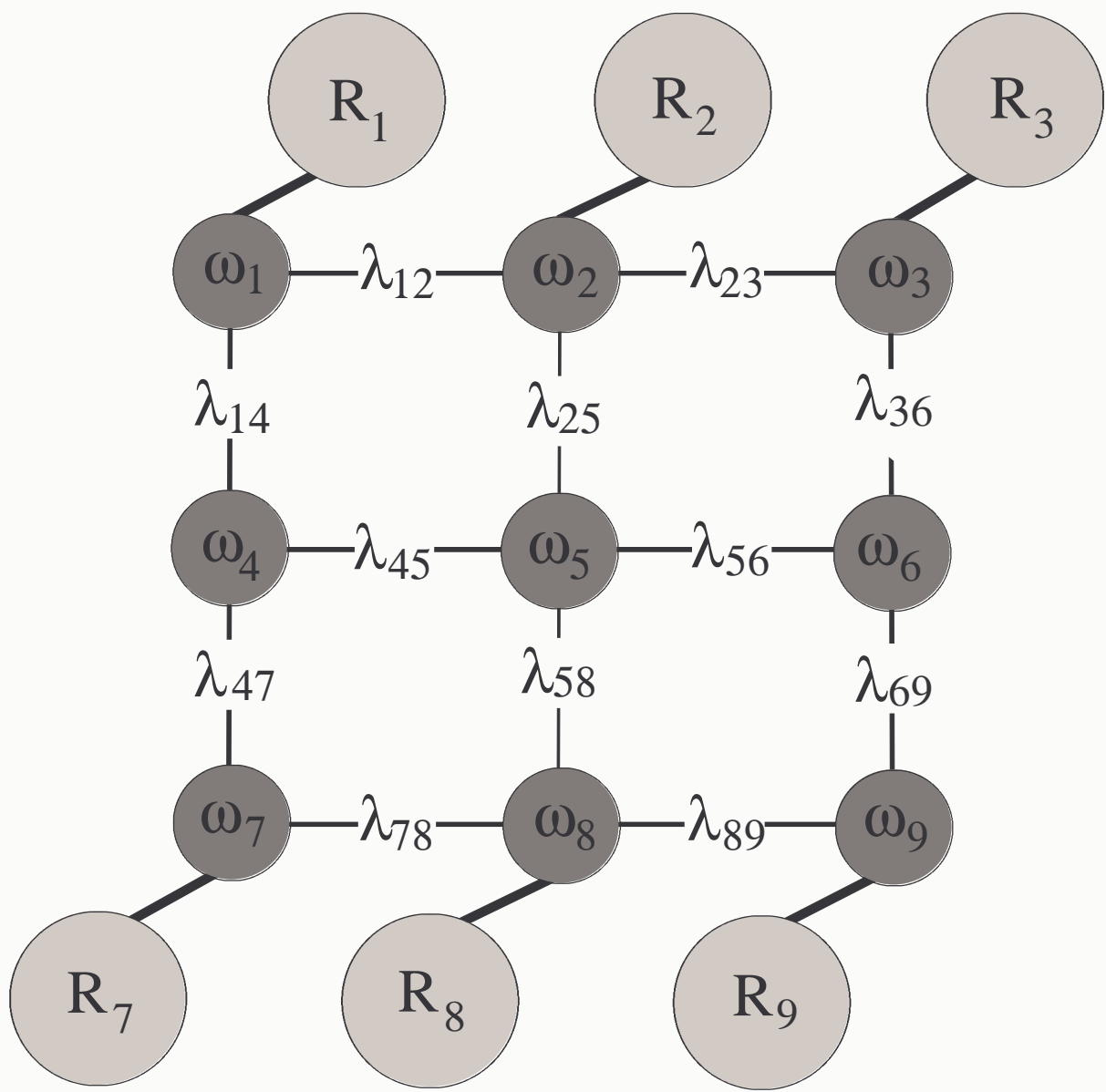
\end{document}